\documentstyle[preprint,aps]{revtex}

\begin{document}

\draft

\title{Atomic Tunneling from a STM/AFM Tip: Dissipative Quantum Effects %
from Phonons}

\author{Ard A. Louis and James P. Sethna}
\address{Laboratory of Atomic and Solid State Physics,
Cornell University, Ithaca, NY 14853-2501}
\maketitle

\begin{abstract}
  We study the effects of phonons on the tunneling of an atom between two
surfaces.  In contrast to an atom tunneling in the bulk, the phonons couple
very strongly, and qualitatively change the tunneling behavior. This is the
first example of {\it ohmic} coupling from phonons for a two-state system.
We propose an experiment in which an atom tunnels from the tip of an STM,
and show how its behavior would be similar to the Macroscopic Quantum
Coherence behavior predicted for SQUIDS.  The ability to tune and calculate
many parameters would lead to detailed tests of the standard theories.
(For a general intro to this work on the on the World-Wide-Web:
http://www.lassp.cornell.edu. Click on ``Entertaining Science Done Here''
and ``Quantum Tunneling of Atoms'')

\end{abstract}

\vspace{1cm}

\narrowtext

The fascinating experiments carried out by Eigler and
colleagues\cite{Eigl90,Eigl91,Stro91,Crom93} have shown that scanning
tunneling microscopy can be used not only for imaging, but also to directly
demonstrate fundamental aspects of quantum mechanics.  Inspired by their
success, we propose an experiment to test
 the effect of a dissipative environment on a quantum mechanical system.
While often ignored in many  applications of quantum mechanics, the environment
can have important effects on a simple quantum system.  In fact,
from the dawn of quantum mechanics, people have been attributing the
collapse of the wavefunction to the interaction of a quantum system with
a macroscopic environment\cite{Zeh70}.
  As recently emphasized by Leggett\cite{Legg92},
condensed--matter physics provides a natural laboratory for studying
this: how does the surrounding macroscopic crystalline environment
change the behavior of an embedded quantum system?  In particular, what can
realistic models of the environment tell us?

A tantalizing view of the importance of the environment is given by the
``overlap catastrophe''\cite{Ande67}. In a metal, the conduction electrons
must be rearranged when the quantum subsystem changes state: the initial and
final electron ground--state wavefunctions are so different that the bath of
conduction electrons can force the subsystem to dissipate energy during each
transition, or even keep the transition from occurring.  In this letter, we
show that phonons, an environment even more ubiquitous than conduction
electrons, can also have an overlap catastrophe.  In particular, we
show that quantum tunneling behavior between two surfaces
 can be dramatically altered by the coupling to phonons.
 We show how detailed calculations of the macroscopic quantum
coherence (MQC) community \cite{Legg87} could be tested
experimentally in an admittedly microscopic system
 using the tunneling of atoms onto and off of the tip of a
scanning tunneling microscope (STM) or atomic force microscope (AFM).

%Eigler and collaborators report on ``transfer near contact'' processes
%\cite{Eigl91,Stro91} in which an STM tip is brought so close to a surface
%that the absorbed Xe atom spontaneously hops to it, a process they attribute
% to thermal hopping.  Intrigued, we pose
% the question - {\em Can the atom also tunnel?}

Eigler and collaborators report on an ``atomic switch'', in which a Xe atom
is reversibly transferred from a Ni(110) surface to the W tip, as well as on
``transfer near contact'' processes in which an STM tip is brought so close
to a surface that the absorbed Xe atom spontaneously hops to
it\cite{Eigl91,Stro91,Lyo91}.  Several theories have been put forth to
explain the atomic switch
experiment\cite{Gao92,Walk93,Bran94,Saen93,Flor93}.  The potentials
calculated for the Xe-tip-surface system have a double well shape as
depicted in figure 1, and bringing the tip just slightly closer to the
surface than the switch operating conditions can give significant tunneling
amplitudes, even for a heavy atom like Xe.  If we take for example some
calculated parameters from reference \cite{Flor93} for Xe on Ni(110) $(V_0 =
7 meV, Q_0 = 0.6 \AA)$, we can obtain
 tunneling elements up to $\hbar \Delta /k_b \sim 0.5 K$! In this paper,
we use the example of Xe simply because a lot is know about that system.
 Lighter
elements such as He would have even larger tunneling amplitudes.
By adding
an electric field, the wells can be biased in either direction quite
sensitively  as
was shown in reference (\cite{Walk93}),
so that by varying the tip position and the bias, both $\Delta$ and $\epsilon$
can be varied independently.
  Of course the exponential dependence of the tunneling
amplitude $\hbar \Delta$ on system parameters and the precise shape of the
potential makes it very difficult to accurately predict its
magnitude.  The point is more that because the atom can be
 brought arbitrarily close to where it sticks to the surface, the tunneling
rate can be made arbitrarily large.

%To model this process, assume the atom is moving in a double well potential
%as shown
%in
%fig(\ref{doublewell}).  For now, we will assume the symmetric form:
%\begin{equation}\label{eq1}
%V(Q) = V_0 \left( 1 - \left( \frac{2 Q}{Q_0} \right)^2 \right)^2.
%\end{equation}
%The zero temperature tunneling amplitude  $\hbar  \Delta$ can be found
%by WKB or an instanton analysis:
%\begin{equation}\label{eq2}
% \hbar \Delta = \left( 1.95 eV.  \right) \left(
%\frac{V_0^3}{Q_0^2 M} \right)^{1/4} \exp [ - 14.7 \sqrt{M V_0 Q_0^2}].
%\end{equation}
%The mass $M$, effective distance $Q_0$, and barrier height $V_0$ are
%in natural units, namely $a.m.u.$, \AA, and $eV$ respectively.

%For atoms such as Xe and parameters suggested by by Stroscio and
%Eiger\cite{Stro91}, we obtain a negligible tunneling frequency due to the
%heavy exponential suppression in (\ref{eq2}).  However, if a significantly
%lighter atom, say He, is used, and we assume $V_0 = 0.25 eV$ and $Q =
%1.2$\AA, we already obtain a frequency of 15 MHz.

We now proceed to the main part of this letter:  the effect of the
phonon environment on our quantum tunneling  system, and
 consider temperatures such
that $k_B T << \hbar \omega_0/2 < V_0$, allowing us to truncate  the Hilbert
 space to two states, one for each well\cite{Dors86}.

As depicted in figure (\ref{atom}), the AFM/STM $+$ atom system will exert a
different force $\pm \Delta{\bf F} /2$ on the surface depending on
where the atom is. The phonons will relax, switching their equilibrium
positions in response to the atom being on the tip or the surface. This can
be modeled by the Hamiltonian of the two-state system with the displaced
phonons:
\begin{equation}\label{eq3}
H = \frac{\hbar \Delta}{2} {\bf \sigma_x} + \frac{\epsilon}{2} {\bf \sigma_z}
  + \sum_{{\bf k}\sigma} \left[ \frac{1}{2} m_{{\bf k}\sigma} x_{{\bf
k}\sigma}^2 {\bf 1}
  + \frac{1}{2} m_{{\bf k}\sigma} \omega_{{\bf k}\sigma}^2
    \left( x_{{\bf k}\sigma} {\bf 1} + q_{{\bf k}\sigma} {\bf \sigma_x}
    \right)^2 \right].
\end{equation}
The $\{x_{{\bf k}\sigma}\}$ are the normal coordinates for a given
polarization $\sigma$ and wavevector ${\bf k}$.  The masses of the normal
coordinate particles are denoted by $m_{{\bf k}\sigma}$, and the phonon
angular frequency is $\omega_{{\bf k}\sigma}$.
The $\{q_{{\bf k}\sigma}\}$
are the new equilibrium positions in the presence of the force
 $\pm \Delta {\bf F} / 2$.

% and for the case of an infinite isotropic
%material we obtain:
%\begin{equation}\label{eq5}
%q_{{\bf k}i} =  \frac{1}{\mu k^2}\left( \delta_{ij} - \frac{(\lambda + \mu)
%q \hat{k}_i\hat{k}_j}{(\lambda + 2 \mu)} \right)
%\frac{1}{\sqrt{N}}
%\sum_l \frac{ \Delta F_j({\bf r}_l)}{2}  e^{- i {\bf k \cdot r}_l},
%\end{equation}
%where $\lambda$ and $\mu$ are the lam\'{e} coefficients,
% and can be determined by solving:
%\begin{equation}\label{eq4}
% q_{{\bf k}\sigma} = D^{-1}({\bf k}) \frac{1}{\sqrt{N}} \sum_l
%\frac{ \Delta {\bf  F}({\bf r}_l)}{2} e^{- i {\bf k \cdot r}_l}.
%\end{equation}
%and the  $\{{\bf r}_l\}$  denote the equilibrium lattice positions of the
%atoms.
% and $D$ is the usual dynamic matrix.  In an isotropic material for
%example, we
%obtain:

In the absence of phonons, the probability of finding the atom at the tip at
time t, given that it was started there at time $t = 0$, oscillates back and
forth with the usual form $P(t) = \sin^2(\Delta t)$. What is now the effect
of the phonons (environment) on the behavior of the atom (embedded quantum
system)?  Is there a {\em qualitative} change of behavior?  For atoms
tunneling between two states in the bulk, an approach often taken to account
for the environment is calculating the overlap integral of the atom in state
one $+ $ relaxed phonons with the atom in state two $+$ relaxed phonons. The
new renormalized tunneling amplitude is just the bare amplitude multiplied
by a Frank-Condon phonon-overlap factor.  The atom still oscillates back and
forth, but now with a reduced
rate.  For atoms tunneling in the bulk, this approach gives good qualitative
results\cite{Seth81,polaron}.
 However, for an atom tunneling between surfaces, the situation is quite
different, as can be seen by considering the force exerted by the atom on
its environment.  While the defect tunneling in a solid only exerts a {\em
dipole} force, resulting in $q_k \sim k^{-1}$\cite{Seth81}, the external tip
$+$ atom system exerts a {\em monopole} force on the surface, resulting in
$q_k \sim k^{-2}$\cite{qk,unpublished}.  The external case has therefore a
stronger coupling at low frequencies, and a naive calculation of the
Frank-Condon factor gives an infrared divergence, a tell-tale sign that the
adiabatic approximation breaks down.  In fact, the case of tunneling between
surfaces corresponds to ``ohmic'' dissipation in the language of the
Macroscopic Quantum Tunneling (MQT) literature, as opposed to the bulk case,
where the dissipation is of the ``superohmic'' variety\cite{Legg87}.  The
effect of the environment can now be characterized by a dimensionless
coupling parameter $\alpha$ defined as:
\begin{equation}\label{eq8}
\alpha = \frac{\eta Q_0^2}{2 \pi \hbar},
\end{equation} with
 $\eta$ equivalent to the friction coefficient in the macroscopic
limit\cite{Cald83}.
  The tunneling element is renormalized to $\Delta_r =
\Delta(\Delta/\omega_c)^{\alpha/1 - \alpha}$ for $\alpha <1$, and is zero
for $\alpha > 1$\cite{Chak82}.  In other words, for small coupling parameter
$\alpha$, the effect of the phonons is {\em quantitative} only, reducing the
tunneling frequency, while if $\alpha$ crosses 1, the effect is {\em
qualitative}:  there is a transition to no tunneling at all! This
fascinating effect is the phonon analogue of Anderson's ``overlap
catastrophe ''
 in an electron gas\cite{Ande67}.  We emphasize that this would be the first
case in which an overlap catastrophe is caused by phonons\cite{Levi94}.

Having shown that an atom tunneling from an STM/AFM tip to a surface
undergoes ``ohmic dissipation'' from the phonon environment, we come to the
question: Is the dimensionless coupling parameter $\alpha$ in an interesting
regime?  Assuming a point force $\Delta  F/2$ on an semi-infinite
isotropic medium with a linear (Debye) dispersion we find:
\begin{equation}\label{eq10}
\alpha = \frac{1}{64 \pi^2 \hbar \rho c_t^3}G(\sigma)(\Delta F)^2,
\end{equation}
 where $\rho$ is the density in $kg \,\, m^{-3}$, $c_t$ is the
transverse sound velocity, and $G(\sigma)$ is a tabulated function of
Poisson's ratio $\sigma = (3 \lambda + \mu)(6(\lambda + \mu))$, with roughly
equal contributions from the acoustic bulk and Rayleigh modes\cite{alpha}.
  The critical $\Delta {\bf F}$ for which $\alpha = 1$ goes from about $3$
nanoNewtons(nN) for W, to $0.3$ nN for Pb.  Another order of magnitude
reduction can be obtained for some organic materials.   At the short
distances we need for atomic tunneling, the typical force of a tip on a
surface is of the order of a nN per \AA \, separation\cite{Duri86},
putting us right into the interesting regime\cite{tipsurf}!

We envisage a number of realizations of the atom-surface tunneling
experiments.
 First, it has been calculated that
for $k_b T \geq \hbar \Delta / \pi \alpha$ the coherent oscillations are
destroyed\cite{Garg85,Mak91}, and   there is only incoherent tunneling with a
rate\cite{Legg87}:
\begin{equation}\label{eq9}
\tau^{-1} = \Delta_r \frac{\Gamma (\alpha)}{\Gamma (\alpha +
1/2)} \frac{(\pi)^{2 \alpha + 1/2}}{2} \left( \frac{k_B T}{\hbar \Delta_r}
\right)^{2 \alpha - 1},
\end{equation}
 from which the parameter $\alpha$ could be extracted and compared to our
predictions.
(For example, an AFM
operating in non-contact mode can measure the oscillations by the force
modulation of $\pm \Delta {\bf F}/2$. The tunneling current of an STM is
significantly larger when the atom is on the tip than when it is on the
surface\cite{Eigl91}.)
 Beautiful experiments\cite{Gold92}
 on  individual two-state tunneling defects
coupled to
conduction electrons in mesoscopic metals, and on SQUIDS\cite{Luke91}
have confirmed this powerlaw behavior of the rate with temperature. However,
in contrast to our system, these experiments do not allow one to vary
$\alpha$, or calculate it from first principles.
  We believe this would be the first quantitative
test of the $\alpha$ parameter, giving valuable insight into the validity of
the linear coupling model of a dissipative environment.
By varying the distance between the tip and the surface, or by varying the
position within the surface unit cell,  the force $\Delta
{\bf F}$, and thus the coupling parameter $\alpha$, could be tuned.  Of
course simultaneously, this would heavily affect the tunneling frequency
$\Delta$ by changing the potential barrier.  However, one could still see the
qualitative change in temperature dependence of the rate.  Of particular
interest would be the crossover from {\em decreasing} with increasing
temperature to {\em increasing} with increasing temperature of the rate when
$\alpha$ crosses  $1/2$ from below\cite{qsr}.

%At high temperatures we predict Arrhenius behavior\cite{Stro91}.  As the
%temperature is lowered, there would be a crossover to power law behavior
%of the rate: $\tau ^{-1} \sim T^{2 \alpha -1}$

Second, a more ambitious experiment might test the coherence predictions:
watching the tunneling turn off as $\alpha$ is increased.  Coherent
oscillations, one of the goals of the MQC community\cite{Legg87}, will occur
if $T < \hbar \Delta /k_B \pi \alpha$, and $\alpha < 1/2$\cite{incoherent}.
 There has been an extensive discussion of the problems of measuring
coherent oscillations in the MQC literature\cite{measurement}.
 For example, Peres\cite{Pere88}
has claimed that non-invasive measurements are impossible in this case: a
measurement introduces and energy of order $\hbar \omega_0$.  This has been
disputed by Leggett and Garg\cite{Legg89}, and in fact, we find that the
uncertainty in momentum $P$ caused by measuring the position to better than
$Q_0/2$ is $(\Delta P)^2/2 \geq \hbar \omega_0 (\hbar
\omega_0/64 V_0)$
 which gives a sizable window for the direct measurement we
propose.  This expression can be easily derived from the
position-momentum uncertainty principle. Peres makes too crude an
approximation for $\Delta P$, and thus overestimates the effect of
localizing the object to be measured (flux in his case) on it's conjugate
variable\cite{unpublished}.
More specifically for the case of Xe on Ni(110)
 theories for the switch experiment estimate that for a bias voltage greater
than $\hbar \omega_0$, about one in $2000$ electrons will inelastically
scatter and excite the atom into a higher vibrational state\cite{Walk93}. For
a current of $\sim$ $0.1 nA$ the atom would be excited on average about once
every $\mu s$.  Thus an STM in imaging mode can only measure oscillations if
$1/\Delta << 1 \mu s$, a rate that is easily reached for close enough
tip-surface separation.  If the bias is less than $\hbar
\omega_0$, excitations can only occur through multiple electron transitions,
so the excitation rate will be much smaller and lower oscillation rates can
be measured.

For experimentally accessible temperatures (say $1 K$) with $\alpha = 0.1$
this means a tunneling rate of at least $10 GHz$.  Recent developments in
high-speed STM have achieved a time resolution of picoseconds\cite{Weis93},
and we envisage a direct measurement of the correlations by sending in pairs
of voltage pulses\cite{measurement}.  This could be accomplished by a
low-temperature STM coupled to picosecond lasers that generate
the voltage pulses(as was done in Refs.\cite{Weis93}).
  Varying the intra-pair time by $\pi / 2 \Delta$ will
give an oscillation of the integrated
 current that reaches a maximum of $40 \%$ of the current difference between
the atom on the tip and the surface for a slightly biased well with
$\epsilon = \Delta$\cite{unpublished}. The interpair time must be much
greater than the intra-pair time, and the pulses must be long compared to the
electron tunneling time (which is on the order of $fs$).
 These would be the first measurements of the coherent oscillations of a
single entity with ohmic dissipation, and while the atom is not
``macroscopic'', our proposed experiment should be a fertile testing ground
of new measurement schemes, and may shed light on the even more difficult
problem of observing MQC in SQUIDS\cite{Wipf87}.

In summary, we have shown how the  phonons couple in a fundamentally different
way to particles at a surface than to particles in a bulk, and can cause an
``overlap catastrophe'', just as electrons can.  More precisely, the coupling
of an atom to phonons at the
surface produces ohmic dissipation, and {\em qualitatively} changes the
tunneling behavior, in contrast to the well-known case of tunneling in the
bulk where the effect of phonons is only {\em quantitative}.  We have
proposed an
experiment with an atom tunneling between an STM/AFM tip and a surface, and
show how this could test theories of MQC, albeit in a microscopic setting.
 Because the parameters $\Delta$, $\epsilon$ and $\alpha$ can be varied by
changing the tip-surface position and the bias, many different regimes of
the theories of
 two-state systems
with ohmic dissipation can be put to the test.

We thank Thom\'{a}s Arias, Sue Coppersmith, Mike Crommie, Shiwu Gao, and
J\"{o}rg Dr\~{a}ger, for encouragement and discussions, Mark Stiles for
pointing out to us the importance of the Rayleigh phonons in our calculation
of $\alpha$, and Mark Stiles and Risto Nieminen for explaining phonon
coupling in atomic scattering to JPS.

This work was supported by the NSF under Grant No. DMR-19-18065

\begin{figure}
\caption[fig1]{{\bf Double-Well potential for the tunneling atom}}
 The atom tunnels from the STM/AFM tip to the substrate surface and back
under the influence of a potential like the one shown above.
For temperatures: $k_BT <<
\hbar \omega_0/2 < V_0$, the tunneling is between the  lowest harmonic
oscillator type states in each well.
 The tunneling
amplitude $\hbar \Delta$ gives the energy splitting of the symmetric and
anti-symmetric superpositions of tip + surface states, and is determined
by the barrier height $V_0$, the effective distance $Q_0$, the mass of
the particle,  as well
as the possible asymmetry energy $\epsilon$. The small oscillation frequency
$\omega_0$ is determined by the same parameters.
\label{doublewell}
\end{figure}

\begin{figure}
\caption[fig2]{{\bf Schematic drawing of the STM/AFM + atom system}}
 The tunneling of the atom couples to phonons via the response of the
substrate to the external force $\pm \Delta {\bf F}/2$ that depends on the
position of the atom. The coupling to an environment drastically alters the
tunneling behavior.   For small coupling and low temperature, the tunneling
is coherent with a renormalized tunneling frequency.  For larger coupling,
and zero temperature, there is no tunneling at all, while at finite temperature
there is incoherent tunneling with a power-law dependence on temperature.
\label{atom}
\end{figure}


\begin{references}
\bibitem{polaron}  A similar effect is at work in the small-polaron
problem, where the phonon-overlap integral slows down the motion of an
intinerant electron.
\bibitem{qk} This can be seen by the following argument:
 $q_{{\bf k}i} \sim D(k)^{-1}_{ij} \sum_l \Delta F_j({\bf r}_l) e^{- i {\bf
k \cdot r}_l}$\cite{unpublished}, where $D(k)_{ij}$ is the dynamical matrix
which goes as
$\sim k^2$ for small k.  Thus $q_k \sim k^{-1}$ for a dipole force, and $q_k
\sim k^{-2}$ for a monopole force.
\bibitem{incoherent} Although the transition to no tunneling only occurs for
coupling $\alpha > 1$, the tunneling in the region $1/2 \leq \alpha < 1$ is
expected to be incoherent\cite{Legg87}.
\end{references}
\end{document}